\documentclass[twocolumn,showpacs,amsmath,amssymb,prd,nofootinbib]{revtex4}
\usepackage{epsfig}
\def\sss{\scriptscriptstyle}
\def\^#1{^{\sss #1}}
\def\_#1{_{\sss #1}}
\def\beq{\begin{equation}}
\def\eeqno#1{\label{#1}\end{equation}}

\def\cmss{{\rm cm~s^{-2}}}

\def\az{a\_{0}}

\def\l0{\ell\_{0}}

\def\rar{\rightarrow}
\def\s{\sigma}
\def\l{\lambda}

\def\f{\phi}

\def\z{\zeta}

\def\r{\rho}

\def\m{\mu}
\def\n{\nu}

\def\A{\mathcal{A}}
\def\B{\mathcal{B}}

\def\G{\mathcal{G}}

\def\R{\mathcal{R}}

\def\b{\beta}
\def\c{\gamma}

\def\vr{{\bf r}}

\def\vR{{\bf R}}

\def\vv{{\bf v}}

\def\vg{{\bf g}}

\def\va{{\bf a}}

\def\vgN{\vg\_N}

\def\grad{\vec\nabla}

\def\gf{\grad\phi}

\def\azg{\A_0}
\def\vinf{V\_\infty}
\def\qz{q\_0}
\def\Qz{Q_0}
\def\bQz{\bar Q_0}
\def\gN{g\_N}

\begin{document}
\title{The road to MOND--a novel perspective}
\author{Mordehai Milgrom }
\affiliation{Department of Particle Physics and Astrophysics, Weizmann Institute}

\begin{abstract}
Accepting that galactic mass discrepancies are due to modified dynamics, I show why it is specifically the MOND paradigm that is pointed to cogently. MOND is thus discussed here as a special case of a larger class of modified dynamics theories whereby galactic systems with large mass discrepancies are described by scale-invariant dynamics.
This is a novel presentation that uses more recent, after-the-fact insights and data (largely predicted beforehand by MOND).
Starting from a purist set of tenets, I follow the path that leads specifically to the MOND basic tenets. The main signposts are: (i) Space-time scale invariance underlies the dynamics of systems with large mass discrepancies. (ii) In these dynamics, $G$ must be replaced by a single ``scale-invariant'' gravitational constant, $\Qz$ (in MOND, $\Qz=\azg=G\az$, where $\az$ is MOND's acceleration constant). (iii) Universality of free fall points to the constant $\qz\equiv \Qz/G$ as the boundary between the $G$-controlled, standard dynamics, and the $\Qz$-controlled, scale-invariant dynamics  (in MOND, $\qz=\az$). (iv) Data clinches the case for $\qz$ being an acceleration (MOND).

\end{abstract}
\maketitle
\section{\label{introduction} Introduction}
MOND \cite{milgrom83,milgrom83a} is a paradigm of dynamics that attributes the mass discrepancies in galactic systems not to dark matter but to a major departure from Newtonian dynamics and general relativity at low accelerations. The basic tenets of the MOND paradigm pertain to the weak-field limit (in the relativistic sense; potential $\ll c$), and are: (i) The modified dynamics involve a new constant with the dimensions of acceleration, $\az$. (ii) Dynamics go to standard dynamics when system accelerations are much larger than $\az$ (formally, in the limit $\az\rar 0$). (iii) Dynamics become scale invariant (SI) for accelerations much below $\az$, which characterize the dynamics of systems that exhibit large mass discrepancies. There, Newton's constant, $G$, which obstructs SI, has to be replaced by the ``SI'' constant, $\azg\equiv \az G$, as the ``gravitational constant''. Formally, this limit is gotten by $\az\rar\infty$, $G\rar 0$, with $\azg$ fixed.
References \cite{fm12,milgrom14c} are recent reviews of MOND.
\par
My purpose here is to show, systematically, how one is lead to these basic tenets, almost inescapably, by the data on galactic mass discrepancies, once one attributes these discrepancies to modified dynamics.
\par
In Sec. \ref{STSI}, I explain why we should accept that galactic systems with large mass discrepancies are governed by SI dynamics. Some general consequences of SI in systems governed by gravity are discuss in Sec. \ref{SIcons}. Section \ref{umb} discusses the transition from the modified regime to the standard regime. In Sec. \ref{WM}, I explain why of all the possible SI dynamics, it is MOND--with the pivotal role of accelerations--that is pinpointed. Section \ref{summ} is a summary and discussion.
\section{\label{STSI}Space-time scale invariance underlies large-mass-discrepancy dynamics}
We start with the observation that the circular rotational speeds around galaxies become asymptotically flat. This is evinced, in the first instance, by the measured extended rotation curves of discs galaxies (see, e.g., Figs. 15, 21-27 of  Ref. \cite{fm12}, and the figures in Ref. \cite{sn07}). The asymptotic-flatness region occurs at large enough radii where the baryonic mass may already be taken as a point mass, and where we thus expect a Keplerian fall-off of the rotational speeds.

\par
Asymptotic flatness of rotation curves--with its clear conflict with the prediction of standard dynamics without dark matter (Kepler's 3rd law)--is, arguably, the most clear-cut and most iconic evidence for a mass discrepancy in galactic systems. These asymptotic regions (and others) are thus characterized by large mass discrepancies.
\par
In advancing MOND \cite{milgrom83}, I have elevated this observation to a cornerstone and an axiom in constructing modified-dynamics explanations of the mass discrepancies: In the thought-for modified-dynamics theory the circular orbital speed around an isolated, bounded mass becomes asymptotically constant, $V(r)\rar\vinf$.
\par
Rotation curves pertain to circular motions of massive test particle. Asymptotic flatness tells us that circular-orbit periods, around a given mass, become proportional to the orbital radius at radii where the mass discrepancy is large. In other words, at such radii all circular orbits around a given mass are related to each other by a scaling transformation
\beq(\vr,t)\rar\l(\vr,t). \eeqno{xxx}
In yet other words, the family of asymptotic, circular orbits around a central body is space-time SI.\footnote{It is assumed that the asymptotic dynamics depends only on the total mass of the central object and not on its size attributes, which also change under scaling.}
\par
There is also evidence that such a space-time scaling applies to other types of orbits in the regions of large discrepancies.
(i) The data and analysis of Refs. \cite{humphrey11,humphrey12} and the subsequent analysis of Ref. \cite{milgrom12}--which concerns extended hot gas in hydrostatic equilibrium around two isolated, elliptical galaxies--is tantamount to showing this SI for the isotropic distribution of velocities in the gas. (ii) The weak-lensing analysis of Ref. \cite{brim13} (see also complementary analysis in Ref. \cite{milgrom13}) extends the evidence for SI to unbound photon trajectories. The analysis of Ref. \cite{brim13}, while consistent with other behaviors, is also consistent with  the gravitational-light-bending angle becoming asymptotically independent of the impact parameter (for radii where the central galaxy is dominant).\footnote{This is equivalent to the finding of the above reference that the observed light bending is consistent with a logarithmic potential.} This implies that asymptotic null-geodesics can be gotten from each other by space-time scaling.
\par
All these would follow if the modified dynamics that describe phenomena exhibiting large mass discrepancies are SI. By this one means that if $\vr_i(t)$ are the solution trajectories, for a self gravitating system of masses $m_i$, for some initial conditions, then, $\l\vr_i(t/\l)$ is the solution for the appropriately scaled initial conditions.
\par
A poor-man's general statement of scale invariance, in the present context, is that velocities--which are invariant to scaling--should depend only on masses, which are invariant, but not on sizes, which are not. This is manifested in the asymptotic flatness of rotation curves, and would also underlie the strong mass-velocity correlations that are known to hold in galactic systems, such as the Tully-Fisher and the Faber-Jackson correlations.
\par
Indeed, many years after the advent of MOND it was realized \cite{milgrom09a,milgrom14} that
this symmetry can be taken as a basic tenets for the modified dynamics, and  many of the observations
that concern phenomena in regions of large mass discrepancies automatically follow (without dark matter). But MOND posits more than SI in large-discrepancy systems; it also pinpoints the special role of accelerations. It is then left to be shown how this is forced on us.
\par
I thus take here the SI in the limiting theory describing high-discrepancy systems as the starting point for constructing the modified dynamics.\footnote{By SI I mean only that the equations of motion are SI, not necessarily the action. For that it is enough that the action has a well-defined scaling dimensions, namely, that it is multiplied by some constant (a power of $\l$) under scaling.} Call this limiting theory the DML theory.\footnote{Since in the end this theory will be identified with the deep-MOND-limit.}
\par
As always in such generalizations, the principle is much more general than the observations on which it is based.

\section{\label{SIcons}Consequences of scale invariance}
We do not know whether the modification of dynamics invoked to explain the discrepancies in the gravity-governed galactic systems has bearings also on non-gravitational physics, such as electromagnetism. Be the case as may be, here I confine myself to the purely gravitational sector of such a modified dynamics.
Furthermore, I confine myself to the nonrelativistic (NR) limit. In addition, and in harmony with this, I assume that in the modified dynamics, bodies (such as galaxies) can still be assigned the attribute of a mass, which is the {\it only} source of their gravity, and which coincides with the standard notion of mass.\footnote{In principle, the modified dynamics could be such that gravity depends on attributes of the sources other than mass.} This mass may have to be measured by means outside the DML theory, for example, its value may be assigned relative to atomic masses or to the Planck mass. Or, it may be taken as the sum of the masses of its constituents (atoms, stars, etc.), which can be determined with standard Newtonian gravity. \footnote{When we say in the context of the mass-discrepancy problem that the baryonic mass falls very short of the measured dynamical mass, we mean by ``baryonic mass'' the sum of the masses of the system's constituents, measured individually by various means. For example, we measure the baryonic mass in neutral hydrogen, by measuring the 21-cm luminosity, which is thought to be proportional to the number of hydrogen atoms, which we multiply by the laboratory value of the hydrogen mass. The baryonic mass in stars is based on masses of individual stars based, in turn, on stellar-evolution models or measurements in stellar binaries, assuming Newtonian dynamics.}
\par
A SI theory can be recognized by its constants; i.e., by whether they retain their values under a simultaneous change of the length and time units by the same factor. More precisely, the degrees of freedom of a theory can be normalized to have units such that this becomes a necessary and sufficient condition for SI.\footnote{When we speak of SI, we are interested in what happens when we scale all lengths and times by a factor $\l$. However, a theory usually involves other degrees of freedom, such as the gravitational potential, that are not directly measurable, call them $\psi$ collectively. To these we can assign arbitrary scaling dimensions, $\xi$, such that it is decreed that under the scaling of eq.(\ref{xxx}), $\psi(\vr,t)\rar \l^{\xi}\psi(\vr/\l,t/\l)$.
By multiplying $\psi$ by powers of the constants, we can always normalize them so that their $[l][t][m]$ dimensions match their scaling dimension. Namely, that if $[\psi]=[l]^\b[t]^\c[m]^\z$, then its $\xi=\b+\c$. With this choice of dimensions, simultaneous change of the length and time units by the same factor, under which all equations are invariant, is equivalent to a scaling transformation (which acts only on the degrees of freedom) followed by the change in the values of the constants. Thus SI is equivalent to invariance of the constants under the unit change.  See more details in Ref. \cite{milgrom14}.} In other words, with this standardized choice of units for the degrees of freedom, which I assume henceforth with no loss of generality, the constants must have dimensions of the form $[m]^\b[\ell]^\c[t]^{-\c}$.
\par
For example, the fact that Newtonian dynamics (and general relativity)--summarized by $\va=-\gf$, $\Delta\f=4\pi G\r$ ($\f$ has zero scaling dimension), or by $a\sim MG/r^2$--is not SI, is manifest by the appearance of $G$, whose value does change under the said change of units.\footnote{In standard units. Equivalently, we can work with units in which $G\equiv 1$, but then masses are not invariant.} These theories can also not be modified into a SI theory without replacing $G$ with other constants that are invariant to the change of units.
\par
In principle, several constants with different dimensions  $\b,~\c$ may appear in such a theory.
In this first decision fork we assume,
for parsimony, that only one such dimensioned constant appears. Such parsimony is not a mere whim. Reason advocates, and historical precedence with relativity and quantum theory concurs, that a major breakdown in well established physics does not introduce more than one new physical constant.
\par
Thus, $G$ is banished from the DML, and replaced with another ``gravitational constant'', call it $\Qz$ (dimensionless parameters may appear, in principle).
\par
We saw that SI predicts the asymptotic constancy of the rotational speed around an isolated body. Now that we accept that only one constant appears in the DML theory, the dependence of $\vinf$ on the body's mass is determined as follows: Since $\vinf$ can depend only on the central mass, we must have, on dimensional grounds,
\beq \vinf\propto (\Qz M^{-\b})^{1/\c}.\eeqno{i}
For $\b=0$, namely, when $\Qz$ has dimensions of velocity to some power, we get that $\vinf$ does not depend on the mass and is a universal constant. This can be rejected at the outset because it clearly conflicts with observations.
\par
Since $\b\not= 0$ we can assume, without loss of generality, that $\b=-1$, which will be our choice of convenience henceforth.\footnote{Starting with some $\bQz$ that has some $\bar\c$, and any $\bar\b\not = 0$, we could choose to work, instead, with $\Qz=\bQz^{-1/\bar\b}$ as our DML gravitational constant. This would have indeed set $\bar\b\rar\b=-1$ for $\Qz$ (and change $\bar\c$ to $\c=-\bar\c/\bar\b$).}
With this standardized choice of units for $\Qz$ we get $\vinf\propto (\Qz M)^{1/\c}$. It remains to fix the numerical normalization of $\Qz$. We take it so that
\beq \vinf= (\Qz M)^{1/\c}. \eeqno{ii}
(In MOND, $\Qz=\azg$.)
This mass-asymptotic-speed relation, $M\propto\vinf^\c$ is an important prediction of SI plus the assumption of only one new constant.
\par
The solitary appearance of $\Qz$ also leads to relations between different DML systems: An immediate corollary is that every NR, DML system is a member of a two-parameter family of homologous systems of different masses and sizes, with a unique relation between masses and velocities, independent of the size.
\par
To see this, start with any NR, self-gravitating, DML system, $S$. Make a change of units of mass, length, and time such that the values of attributes with these dimensions change as: $m\rar m'=\theta m$, $\ell\rar \ell'=\kappa\ell$, and $t\rar t'=\kappa\theta^{-1/\c}t$. The value of $\Qz$ remains the same in the new units, while the values of velocities become
$\vv\rar \vv'=\theta^{1/\c}\vv$. Since all equations are invariant to a change of units, a system homologous to $S$ with the tagged attributes is also a solution of the DML theory (with the same value of $\Qz$). We thus generate a whole two-parameter family of homologous solutions of all masses and sizes, within which all velocities (rotations, dispersions, etc.) are independent of size, and scale as
\beq V\propto (\Qz M)^{1/\c}. \eeqno{iii}
The dimensionless proportionality factor in eq.(\ref{iii}) may be different for different families, and may depend on dimensionless attributes, such as the different mass ratios of constituents, velocity anisotropies, shape parameters, etc.\footnote{A different, two-parameter scaling, applies, of course,  in Newtonian dynamics due to the solitary appearance of $G$.}
\par
For example, it follows that all DML-governed thin-disc galaxies with an exponential surface density profile, of all masses, and all exponential scale lengths, have the same value of $V^\c/M$.
\par
Despite the similar appearance, relation (\ref{iii}) is very different from the $M-\vinf$ relation (\ref{ii}). The one concerns asymptotics, the other bulk properties;  the one applies (asymptotically) to all systems, the other only to DML systems; the one predicts a universal equality with no scatter, the other not, as the ratio $V^\c/M$ may depend on dimensionless attributes of the system.

\section{\label{umb}The umbrella theory}
The DML theory, epitomized by the solitary appearance of $\Qz$, is only part of the weak-field physics needed to describe gravitating systems. We know that it can only be a limit of a more complete theory, another limit of which is the weak-field general relativity (or Newtonian dynamics), epitomized by the solitary appearance of $G$. The umbrella theory involves both constants.
\subsection{The boundary constant}
We expect the boundary between the validity domains of these two limits to be defined by some dimensioned constant, as $\hbar$ defines the boundary between the classical and quantum domains, or $c$ the classical-relativity boundary.
Again, being loath to add yet more constants, I assume that this ``boundary constant'' is constructed from $G$ and $\Qz$.
\par
We want to retain the weak equivalence principle, namely the universality of free fall of test particles, in the DML, as well as in the full umbrella theory.\footnote{Some observational evidence for this comes, for example, from the fact that different test bodies, such as gas atoms, stars, etc. show the same rotation curves within observational errors. The bounds on departure from the principle are much looser than in laboratory and solar system measurements.} This dictates that the ``boundary constant'' has no mass dimensions. Otherwise, the question of whether some test particle motions are described by the DML or by Newtonian dynamics will depend on the mass of the particle, not only on the motion.
\par
To see this, consider a test particle of mass $m$ in the system, and let $\B(\psi,G,\Qz)$ be a dimensionless criterion number, whose value tells us whether we are in the DML or not (similar to $h\n/kT$ in connection with the black-body spectrum in the classical-quantum case, or $v/c$ and $MG/rc^2$ in relativity). Here, $\psi$ represents some local and nonlocal characteristic(s) of the particle's orbit. Universality of free fall dictates that $\psi$ cannot depend on $m$. Since $\B$ is dimensionless, it is invariant under a change of the mass units, under which the values of $G$ and $\Qz$ do change. So, only the ratio $\qz=\Qz/G$, whose value does not change (since with our standardized choice of $\Qz$ it has the same mass dimensions as $G$), can appear in $\B$. Furthermore, $\B(\psi,\qz)$ can depend only on $\psi$ that are either dimensionless or of the dimensions of $\qz$ and then appear as $\psi/\qz$.
\par
This is the case with $c$ as the classical-relativistic ``boundary constant'', which does not break universality of free fall. On the other hand, $\hbar$, as the classical-quantum ``boundary constant'', does have mass dimensions, leading to manifest breakdown of the universality of free fall: For a given motion (velocity), the de Broglie wavelength of a particle depends on its mass.\footnote{So, for example, the analogue of the Bohr radius of a test mass $m$ in the gravitational field of a finite mass $M$, $r\_B\sim \hbar^2/m^2MG$, depends on $m$. (However, relative to the Planck length, $r_p$, we have $r\_B/r_p\sim M_p^3/m^2M$, where $M_p$ is the Planck mass.)}
The appearance of $G$ in Newtonian dynamics is not an obstacle, as it is not a boundary constant, only a conversion constant between inertial and gravitational masses.
\par
The boundary can thus be defined by $\qz$, but it may depend on dimensionless characteristics of the orbit or phenomenon in question, such as the type of orbit (e.g., whether it is circular or not, etc.). So, for certain phenomena the actual boundary may, in principle, be a large or small multiple of $\qz$.
I assume that this does not happen, namely, the constant that emerges from the DML's value of $\Qz$ (as $\qz=\Qz/G$) also marks the approximate boundary for all phenomena. Such is the case, of course, for the transitions from classical physics to quantum physics and to relativity. The last verdict on this assumption is left to the observations.

\par
If all the test particles from which a system is made are always on one side of the boundary, we may say that the whole system is on that side. In this case we may express the characteristic value of $\psi$ for a typical test particle in terms of the global properties of the system (such as mass and size), and get a global criterion for the system to be on that side of the boundary. For example, in the case of MOND where $\qz=\az$, such a system parameter may be $\s^2/R$, or $MG/R^2$, or $(M\azg)^{1/2}/R$. This is in the basis of the statement that a low-surface-density system: with $MG/R^2\ll\az$ is wholly in the DML. Note that here the total mass $M$ does appear; this is not in conflict with the weak equivalence principle.

\par
The dimensions of $\qz$ are
\beq [\qz]=[\ell]^{\c-3}[t]^{-(\c-2)}. \eeqno{x}
In terms of $\qz$, eq. (\ref{ii}) reads
\beq \vinf= (\qz GM)^{1/\c}, \eeqno{v}
and eq.(\ref{iii}) reads
\beq V\propto (\qz GM)^{1/\c}. \eeqno{vi}
For example, if dynamics is modified beyond a certain scale length, $\ell_0$, we have to take $\c=2$, so $\qz\sim\ell_0^{-1}$ is an inverse length. This predicts $\vinf\propto M^{1/2}$.
If we want to modify dynamics below a critical frequency, $\omega_0$, we have to take $\c=3$, so $\qz\sim\omega_0$. In this case $\vinf\propto M^{1/3}$.
The MOND choice is $\c=4$, namely, $\qz=\az$ is an acceleration, which predicts $\vinf\propto M^{1/4}$.

\subsection{A counterexample}
To recapitulate an important point: I have assumed that if the NR theory that I treat all along is attained as the NR limit of some relativistic theory, then this limit is strictly $c\rar\infty$, so that $c$ does not appear beside $G$ in the NR limit. The NR theory is then assumed to have both a Newtonian limit, controlled, as usual, by $G$ only, and a SI limit controlled by only one constant $\Qz$.
\par
As an instructive exception I explain why so-called $F(\R)$ theories, or rather their NR limits, are not of the type discussed here, which disfavors them as candidate alternatives to dark matter.
\par
In such theories, the Einstein-Hilbert Lagrangian $L\propto c^4G^{-1}\R$ ($\R$ is the Ricci scalar), is replaced by $L\propto c^4G^{-1}\ell_0^{-2} F(\ell_0^2 \R)$, where $\ell_0$ is some length constant of the theory.
The strict NR limit, which assumes expansion near a Minkowski space time, is attained by taking $c\rar \infty$, with $\ell_0/c\equiv\tau_0$ and $c^2/G$ fixed. In this limit, $\ell_0^2 \R\propto \tau_0^2\Delta\f$, where $\f$ is the gravitational potential defined by $g\_{00}\equiv -1-2\f/c^2$ (where $g\_{00}$ is the time-time component of the metric). The Lagrangian then goes to
\beq L\_{NR}\propto c^2G^{-1}\tau_0^{-2} F(\tau_0^2 \Delta\f). \eeqno{juna}
But this $L\_{NR}$, which involves both $G$ and $c$, does not have a Newtonian limit for any choice of $F(\chi)$.\footnote{In the relativistic theory, the general-relativity limit is obtained by taking $F(\chi)\propto \chi$, but if we take this limit in the NR Lagrangian, it becomes a complete derivative, and is unacceptable. The two limits--the NR limit and the GR/Newtonian limit--are not interchangeable.} (The particle Lagrangian is assumed intact and does not introduce any constants except masses.)
\par
The NR Lagrangian, $L\_{NR}$, does have a SI limit, which necessitates that in this limit $F(\chi)\propto \chi^{3/2}$, so $L\_{NR}\propto \Qz^{-1}(\Delta\f)^{3/2}$, with $\Qz=G/c\ell_0$, the dimensions of which are $[\Qz]=[m]^{-1}[\ell_0][t]^{-1}$, so we get a theory with $\c=1$. This theory is conformally invariant in the three-dimensional Euclidean space.
\par
This higher-order theory (which suffers from the usual maladies of such theories) demonstrates also the need for our assumption that the asymptotic dynamics depends only on the total mass.
The field equation
\beq \Delta[(\Delta\f)^{1/2}]\propto\Qz\r, \eeqno{buta}
has four independent spherical vacuum solutions: $\f=(\Qz M)^2\times[(r/r_0)^{-1},~ln(r/r_0),~const., ~(r/r_0)^2]$. Only the 2nd gives asymptotically flat RCs.  That is because for the others solutions, the asymptotic dynamics, which is determined by $\gf$, depends not only on the total mass, but also on structural attributes with the dimensions of length, which enter through $r_0$, and which must also be scaled. So, the scaled orbit does not correspond to the same mass system.

\subsection{The discrepancy: definition and predictions}
The discrepancy between the observations and the predictions of standard dynamics without dark matter is usually found and characterized by comparing the measured acceleration, $\vg$, of test particles on some orbits at some location in a gravitating system, with the acceleration, $\vgN$, predicted by standard dynamics assuming that only the observed matter is present.
Thus, as some measure of the discrepancy we can take
$\eta\equiv g/\gN$, where $g=|\vg|$, $\gN=|\vgN|$. The ratio
$\eta$ is sometimes called ``the mass discrepancy''. The heuristic reason for this is clear: For a spherical system, Newtonian dynamics say that at a given radius the acceleration is proportional to the enclosed mass. So $\eta$ would indeed be the ratio between the enclosed mass needed to produce $\vg$ to that producing $\vgN$.
But in general, when the mass distributions of baryons and of the dark matter--real or fictitious--are not spherical, and are different, the acceleration does not generally measure some enclosed mass, so the ``acceleration discrepancy'' $\eta$ is not really a ``mass discrepancy''; the term can only be used loosely. In fact, since in general $\vg$ and $\vgN$ are not parallel, we could define several acceleration ratios, for example using separately the acceleration component ratios.
\par
In this connection, note the following cautionary example: Ref. \cite{loebman14} analyzed the out-of-the-galactic-plane dynamics in the Milky Way. As regards MOND, they considered, separately, the ratios of the radial and perpendicular components of $\vg$ and $\vgN$, call them $\eta_r=g^r/\gN^r$ and $\eta_z=g^z/\gN^z$, respectively. They plot these discrepancy measures against the acceleration. They find that, in accordance with the predictions of MOND, both discrepancies set in roughly below an acceleration of $10^{-8}\cmss$, and that $\eta_r,~\eta_z$ increase with the decrease in $g$ as MOND prescribes. There is no known reason why these clear systematics should happen in the dark-matter paradigm. But Ref. \cite{loebman14} fault MOND since they find that $\eta_r$ and $\eta_z$ are not quite the same.  The MOND expression they used erroneously, $\vgN=\vg\m(g/\az)$, applies only in high-symmetry situation, such as spherical symmetry, or the axisymmetric case relevant for rotation curves. In such cases, clearly, $\vg$ and $\vgN$ must be parallel from symmetry, and so automatically $\eta_r=\eta_z$. However, there is no full-fledged MOND theory that predicts them to be the same for the analysis of Ref. \cite{loebman14}. It is also to be noted that the results of such an analysis depend on the adopted, but rather uncertain, model for the distribution of baryons in the Milky Way, which determines the values of $\vgN(\vR)$.
\par
Furthermore, in Newtonian dynamics, the accelerations of all test particles are the same at a given position; so $\eta$ would be a function of position only. This is also the case in modified-gravity theories where the NR accelerations are still given by the gradient of a potential. But in a subclass of modified-dynamics theories, so termed ``modified inertia'' theories, defined and described in Ref. \cite{milgrom94}, the accelerations may depend not only on location but on the whole orbit. So particles on different orbits might give rise to different values of $\eta$ even if their acceleration is measured at the same position.
\par
Returning to our general discussion,
what do our modified-dynamics theories predict for $\eta$? In general, we expect that $\eta$ should be correlated with the value of system attributes, $q$, that have the dimensions of $\qz$, such that when $q\gg\qz$, $\eta\approx 1$ (no discrepancy), while for $q\ll\az$, $\eta\gg 1$.\footnote{On the asymptotically flat section of rotation curves, the discrepancy increases with radius. Also, a quantity with the dimensions of $\qz$ has dimensions of velocity$^{\c-2}$/length; so must decrease as the inverse radius. Thus, this is indeed the sign of the inequalities.}
\par
When $g$ is determined from the asymptotic parts of rotation curves of disc galaxies, our theories predict
$g=\vinf^2/r$, with $\vinf$ given by eq.(\ref{v}), while $\gN=MG/r^2$, where $M$ is the total (baryonic) mass of the galaxy.
Thus,
\beq \eta=\eta\_{\R}=\frac{\qz}{q\_{\R}(r)}, \eeqno{vii}
where $q\_{\R}(r)\equiv \vinf^{\c-2}/r$ (subscript $\R$ indicates that the quantity pertains to rotation curve).
$q\_{\R}$ is a local parameter that has the dimensions of $\qz$, according to eq. (\ref{x}). The same result is gotten when $\eta$ is obtained from weak lensing by galaxies, when  analyzed consistently assuming asymptotically flat rotation curves \cite{milgrom13}.
\par
More generally, defining $q\_{\R}(r)\equiv V^{\c-2}(r)/r$ for the full rotation curve, we expect $\eta\_{\R}$ to be strongly correlated with, if not quite a function of, $x=q\_{\R}(r)/\qz$. Any such correlation is predicted to give $\eta\approx 1$ for $x\gg 1$, and $\eta=p/x$ for $x\ll 1$, where $p\sim 1$ is a proportionality factor that may depend on the exact phenomenon and orbit, and other aspects entering the measurement of $\eta$. The latter proportionality stems from SI: From their definition, $\eta$ scales as $V^{2-\c}r$, while $x$ scales as $V^2/r$. Since $V$ is invariant, we have under scaling $\eta\rar\l\eta$, while $x\rar x/\l$.
\par
When $x\ll 1$ is achieved on the asymptotic parts of rotation curve we have $p\equiv 1$ due to our normalization of $\Qz$ and $\qz$. But for other instances of $x\ll 1$ only proportionality is predicted. For example, circular, but non-asymptotic orbits, or radial orbits, or the use of different acceleration components to define $\eta$,  might give other values of $p$. These details depend on the theory.
\par
Modified inertia theories of the type discussed in Ref. \cite{milgrom94} do predict a strict functional dependence\footnote{Ref. \cite{milgrom94} deals specifically with MOND, but the same argumentation applies to our more general $\c$ case.}
\beq \eta\_{\R}=f(q\_{\R}/\qz),~~~~f(x\gg 1)\approx 1,~~~~f(x\ll 1)\approx x^{-1}.  \eeqno{xi}
That $f(x\ll 1)\approx x^{-1}$ follows from SI and the standardized definition of $\qz$, since, for the same reasons as above, $\eta$ scales like $\l$, while $q$ (hence $x$) scale as $\l^{-1}$.
This functional dependence does not extend beyond rotation curves; in fact we do not have a general result for such ``modified-inertia'' theories.
\par
Now consider how $\eta$ varies not within a given system, but across different systems. Write some global discrepancy measure, $\eta\_{\G}$, in terms of the global attributes of DML systems: the total mass, $M$, some measure of the radius, $R$ (e.g., the scale length of exponential discs, the half-mass radius, etc.), and some representative measure of the internal speeds, $\s$ (e.g. the mass-weighted root-mean-squared velocity).

\beq \eta\_{\G}\equiv (\s^2/R)/(MG/R^2)=\s^2R/MG.  \eeqno{xiii}
Also, define a global attribute with the dimensions of $\qz$:
\beq q\_{\G}=\s^{\c-2}/R.  \eeqno{xxxii}
The scaling arguments around eq.(\ref{iii}) tell us that within each two-parameter family of systems,
 \beq \eta\_{\G}=\kappa\_{\G}\frac{\qz}{q\_{\G}}, \eeqno{xiv}
where $\kappa\_{\G}$ is a dimensionless constant that might depend on dimensionless attributes of the family, and on the exact definition of our global system attributes (unlike the case of rotation curves where the choice of these parameters is well defined).
\par
For Newtonian systems, which have $q\_{\G}\gg\qz$, the Newtonian virial relation implies $\eta\_{\G}\sim 1$.
\par
Now, to compare between different families of DML systems, we use the assumption of no dimensionless constants that much differ from unity. This tells us that in each family there are members near the MOND-Newtonian boundary where we have both $q\_{\G}\sim\qz$ and $\eta\_{\G}\sim 1$. This means that $\kappa\_{\G}\sim 1$ for all the families.
\par
The important consequence is thus that $\eta\_{\G}$ is strongly correlated with $q\_{\G}$ for all systems, such that $\eta\_{\G}\sim \qz/q\_{\G}$ when $q\_{\G}\ll\qz$, and $\eta\_{\G}\approx 1$ when $q\_{\G}\gg\qz$. In other words, a strong correlation of the form described in eq.(\ref{xi}) is predicted.
\par
For example, if the modification hinges on distance ($\c=2$), introducing as boundary a new length $\ell_0$, we predict a correlation\footnote{This and the ones below are predicted correlations not exact functional relations (hence the use of the $\sim$ sign). Different phenomena, or different kinds of orbits may show somewhat different behaviors.}
\beq \eta\sim f(R/\ell_0). \eeqno{xxii}
In the case of MOND ($\c=4$), we predict a correlation
\beq\eta\sim f(g/\az),\eeqno{xvi}
which can also be written as \cite{milgrom83}
\beq\eta\sim \n(\gN/\az), \eeqno{xv}
because $\eta\equiv g/\gN$.
For $\n(y)$ we have $\n(y\gg 1)\approx 1$, and in the DML, SI requirs $\n(y\ll 1)\approx y^{-1/2}$.

\section{\label{WM}Why an acceleration (why MOND)?}
To complete the road to MOND we have to pinpoint $\c$ and hence the full dimensions of $\qz$ and $\Qz$.
\par
We have identified above at least four regularities in which $\c$ appears and from which it can be determined.
(i) The mass-asymptotic-speed relation eqs.(\ref{ii})(\ref{v}). (ii) The general virial relation between internal velocities and total mass, eqs.(\ref{iii})(\ref{vi}). (iii) The identification of the ``boundary constant'' and its dimensions. (iv) The dimensions of the system attribute with which the discrepancy correlates.

The mass-asymptotic-speed relation is closely related to the observed Tully-Fisher relation, which started life as a correlation between galaxy luminosity in some photometric band, and some measure of the rotational speed (e.g., the 21-cm line width). Many versions of such a relation have been presented in the literature, using various measures of the rotation speed and various luminosity measures. As has been stressed by MOND from its advent, and as is clear from our discussion above, SI modified dynamics require strictly that the velocity measure used is the asymptotic constant speed. It is also required (as emphasized, e.g., in Ref. \cite{mb88}) that the other variable be the total baryonic mass, not a luminosity, which at best measures the stellar mass, while many galaxies can have a substantial fraction of their mass in gas. As a result of these MOND admonitions we have seen more appropriate versions of the Tully-Fisher relation. For example, Ref. \cite{nv07} found that using the asymptotic speed for a sample of high-mass galaxies gave a tighter correlation than using other velocity measures. For the first they found a slope of $0.24\pm0.07$.\footnote{Ref. \cite{nv07} still used a luminosity measure, but it was the $K$-band luminosity, which is thought to be a good measure of the stellar mass. And the gas contribution to the mass is small in the sample studied. Their sample included a relatively small range of galaxy masses; reducing the leverage on the slope.} Reference \cite{mcgaugh11a} tested the predicted MOND baryonic Tully-Fisher relation for galaxies whose mass is dominated by gas (to minimize the effect of light-to-mass conversion for the stars). The result of this analysis, and of one including all types of galaxies, is shown in Fig. 3 of Ref. \cite{milgrom14c}. It was found that the $M-\vinf$ slope is very tightly constrained to be $\c\approx 4$. Slopes of $\c=2$ (large distance modification) and $\c=3$ (low frequency modification) are clearly excluded.\footnote{Even at the time MOND was put forth, it was possible to reject large-distance modifications ($\c=2$) on the basis of existing limits on the Tully-Fisher slope \cite{milgrom83a}.}
\par
Some of the weak-lensing results of Ref. \cite{brim13}, can also be viewed as constraining $\c$: When using logarithmic potentials (equivalent to asymptotically flat rotation curves) to fit their lensing data (binned into several luminosity bins spanning 2.5 orders of magnitude), Ref. \cite{brim13} plot the galaxy luminosity vs. the equivalent of $\vinf$. They analyzed separately red (elliptical) galaxies, and blue (disc) galaxies and find for the respective slopes
$\c^{-1}=0.24\pm 0.03$ and $\c^{-1}=0.23\pm 0.03$.
\par
The general mass-velocity, virial relation for large-discrepancy systems, while well consistent with $\c=4$
(e.g., Refs. \cite{mm13,mm13a}) does not provide as tight a constraint as others due to the caveats mentioned in Sec. \ref{umb}.
\par
To top the argument, Fig. 5 of Ref. \cite{milgrom14c} (produced by Stacy McGaugh)\footnote{This is an up-to-date version superseding previous similar analyses, e.g. in Refs. \cite{sanders90}\cite{tiret09} for disc galaxies, and in Ref. \cite{scarpa06} for pressure-supported systems.}, shows the discrepancy, $\eta\_\R$, as derived from the rotation curves of 73 disc galaxies, at many radii, $r$. $\eta$ is first plotted as function of $r$, to see if a correlation of the kind (\ref{xxii}) exists, which $\c=2$ predicts. No such correlation appears.
\par
Then, this figure shows a test the predictions of the case $\c=4$ (MOND): Equation (\ref{xv}) [and so also of (\ref{xvi})] is tested by plotting $\eta\_\R$ against $\gN$. It was found that, indeed, a strong correlation exists, which is consistent, in fact, with no intrinsic scatter.
\par
In such a figure, $\az$ appears in three different roles relevant to our discussion above: (i) Some of the points in the region $\gN\ll\az$ (or $g\ll\az$) correspond to asymptotic regions of the rotation curves. The fact that they give $\eta\approx (\az/\gN)^{1/2}$ reflects the validity of the prediction of eq. (\ref{v}). (ii) Other points in this region correspond to radii within the bulk of galaxies that show large mass discrepancies everywhere (so called ``low-surface-brightness galaxies''). Here, the more general prediction is vindicated. (iii) The transition from $\eta =1$ to the DML asymptotic regime occurs at $\approx \az$ of the same value. The transition occurs in a span of accelerations of order $\az$ itself (roughly between $\az/2$ and $2\az$). This vindicates our surmise that no new dimensionless constants are introduced, so that $\az$ defined from the DML is also the ``boundary constant'', and the width of the transition region is of the order of $\az$ itself.

\section{\label{summ}Summary and discussion}
I have considered a ``purist'' picture of modified dynamics as an alternative to dark matter. In this, only one new dimensioned constant is introduced by the new dynamics, which might be taken as $\az$, or equivalently as $\azg$. I also assumed that no dimensionless constants that differ much from unity enter the new dynamics. Such was the case in the classical-to-quantum and classical-to-relativistic extensions.

Here again are the main signposts on the road to MOND:

(i) Asymptotic flatness of rotation curves--elevated to a status of an axiom--implies that within a given galaxy orbital times are proportional to the radius.
Among different large-discrepancy objects, velocities depend only on mass not on size; namely, dynamical times are proportional to size. I concluded that it behooves us to adopt SI dynamics for systems showing large mass discrepancies (so termed DML dynamics).

(ii) I assumed that, replacing $G$, a single dimensioned gravitational constant, $\Qz$, appears in the DML (apart from particle masses), which can be generally standardized to have dimensions $[\Qz]=[m]^{-1}[\ell]^\c[t]^{-\c}$.

(iii) I assumed that the Umbrella theory encompassing the DML and Newtonian dynamics, which involves $G$ and $\Qz$, does not involve additional dimensioned constants, nor dimensionless constants that much differ from unity.

(iv) I assumed that the Umbrella theory satisfies the universality of free fall, and hence the constant that marks the boundary between the two limits of the theory has to be (after some standardization) $\qz=\Qz/G$, which has dimensions $[\qz]=[\ell]^{\c-3}[t]^{-(\c-2)}$.

(v) Finally, based on many pieces of evidence I argued that $\c=4$, pinpointing $\Qz=\azg$ as the MOND gravitational constant, and $\qz=\az$ as the MOND acceleration constant.\footnote{Of course we cannot say that the data gives strictly $\c=4$, since it is subject to measurement and systematic uncertainties. But certainly $\c=4$ is by far the most acceptable integer.}
\par
MOND, with its revolving around an acceleration, also turns out to have an added appeal in that the measured value of $\az$ might have strong cosmological connotations (e.g., Ref. \cite{milgrom83}).
For example, $2\pi\az\approx cH_0\approx c^2(\Lambda/3)^{1/2}$, where $H_0$ is the Hubble constant, and $\Lambda$ the ``cosmological constant''. There would be no such a connection had the transition from standard dynamics to the SI dynamics occurred at some critical lengths or time.
\par
The above coincidence also means that the MOND length, $\ell\_M\equiv c^2/\az$, is of the order of the Hubble distance, or the de Sitter radius associated with the cosmological constant. Thus, the role of $\az$ as a ``boundary constant'' can also be expressed as the role of the cosmologically-significant length $\ell\_M$ as a boundary constant: There is a marked transition from non SI dynamics when the length $\ell(a)=c^2/a$, associated with an acceleration $a$, is smaller than $\ell\_M$, to SI dynamics when $\ell\gg\ell\_M$ \cite{milgrom99,milgrom14c}.
\par
While my minimalist and purist starting point may, in the end, prove too restrictive in the case of MOND, I see no cogent reasons, at present, to relax it. On the other hand, there are quite a few theories that purport to underlie MOND and that transgress these axioms, such as hybrid, dark-matter-MOND theories, or those that allow $\az$ to be system dependent (see, e.g., Ref. \cite{milgrom14c} for a review).
\par
Note also, in this connection, that the salient aspects of MOND can be described as a configuration-dependent renormalization of Newton's constant. For DML configurations the main MOND results amount to a renormalization
\beq G\rar \azg/a=(\az/a)G=[\ell(a)/\ell\_M]G. \eeqno{sasum}
A mass discrepancy also appears in the context of the measured expansion history of The Universe after the decoupling of baryons from radiation, a period when NR matter strongly dominates over radiation. As evidenced by several observations, within standard dynamics, baryons alone cannot account for the expansion history. Results are standardly expressed in terms of the baryon and cosmological dark-matter densities. But we can say, alternatively, using the best figures we now have \cite{planck15}, that baryons fall short by a factor of $2\pi$ (to within the uncertainties, which are a few percents). Instead of interpreting this as a $\sim 1:5$ baryon-to-dark-matter ratio, in the spirit of MOND we would note, instead, that a renormalization $G\rar \approx 2\pi G$ would have an equivalent effect.\footnote{Note that ``dark matter'' is also required in standard dynamics to explain structure formation and the microwave background power spectrum.}$^,$\footnote{For much earlier times, Ref. \cite{carroll04} derived a rather tight limit on the value $G'$ of $G$ that enters the expansion history at the time of nucleosynthesis, a few minutes after the big bang, when radiation (relativistic matter) was dominant. From the  observed $^4{\rm He}$ abundance they deduced an upper limit of $|G'/G-1|\lesssim 0.13$.}
\par
The idea of configuration- or experiment-dependent renormalization of $G$ in the basis of MOND is appealing. But, in default of a microscopic theory that leads to such varied results, the idea remains a rather empty truism.
\par
All such, more baroque, attempts at underlying theories come mainly to remedy the shortcoming of MOND in accounting for the cosmological mass discrepancies. But we do not necessarily need to go beyond the purist picture to account for the cosmological anomalies. Relativistic MOND theories that embody the purist tenets (in their NR limits) may still turn out to be rich enough in structure and possibilities to extend the MOND concept to cosmology. Of the presently known such classes of theories we may mention BIMOND (bimetric MOND) \cite{milgrom09}, and nonlocal metric theories \cite{deffayet14}.
\par
There is an illuminating, formal analogy of our discussion in this paper with the phenomenon of propagation of gravity waves in non-viscous fluids, free of surface tension. For a large pool of fluid of constant depth $h$, in a gravitational acceleration field $g$, the dispersion relation is
\beq \omega^2=gk\cdot{\rm tanh}(kh),   \eeqno{grawav}
where $\omega$ is the wave (angular) frequency, and $k=|{\bf k}|$, where ${\bf k}$ is the wave vector.
\par
To underscore the analogy, write eq. (\ref{grawav}) in terms of the following quantities: Instead of the constants $g$ and $h$, use the combinations: $c_s\equiv (gh)^{1/2}$ (a speed), and $\bar G\equiv c_s^2 h=gh^2$, and instead of the variables $\omega$ and $k$, use the variables $r_s\equiv c_s/\omega$, which is a length, and the acceleration $a\equiv c_s^2 k$.
We then have
\beq \bar G/r_s^2=a\cdot{\rm tanh}(a/g).   \eeqno{grawavmod}
This has the form of a basic MOND relation, with the constants $g$, $h$, $\bar G$, and $c_s^4=g\bar G$ analogous, respectively, to  $\az$, $\ell\_M$, $G$, and $\azg$ in MOND, and the variables  $a$ and $\bar G/r_s^2$ the analogs of the acceleration calculated in MOND, and the Newtonian acceleration $G/r^2$ (the gravitating mass is 1 in this analogy).
\par
The Newtonian limit of MOND is analogous to the deep-pool limit, $h\gg k^{-1}$ ($\ell\_M\rar\infty$), which can be affected by taking formally $g\rar 0$ ($\az\rar 0$), with all other quantities fixed. Then dynamics involve only the constant $\bar G$, which is not invariant to unit scaling. As in Newtonian dynamics the resulting relation $\bar G/r_s^2\approx a$ is not SI. The DML corresponds to the opposite limit: $g\rar \infty$  ($\az\rar\infty$), with $c_s$ fixed, so $\bar G\rar 0$ ($G\rar 0$). The dispersion relation then involves only $c_s$ and is SI, as it takes the form $c_s^4/r_s^2\approx a^2$.\footnote{The analogy is even closer in the heuristic picture described in Ref. \cite{milgrom99}, where the wavelength $2\pi k^{-1}$ corresponds to the Unruh wavelength associated with the acceleration $a$.}
\par
This two-dimensional description of the system is clearly only an approximate, effective account of the full three-dimensional hydrodynamical behavior. It is only valid in the limit where the bulk speed of sound of the fluid $c_b\gg c_s$. It not only exemplifies the MOND behaviors in the two limits, but also
demonstrates how an explicit ``interpolating function'' [$\m(x)={\rm tanh}(x)$] between the limits emerges from basis considerations in an effective theory (similar to what may happen in the case of MOND--as has been stressed many times).
This interpolating function has, in fact, the exact asymptotic behaviors, at both ends, as is required in MOND [$\m(x\gg 1)\approx 1$, $\m(x\ll 1)\approx x$].
The system also satisfies our other assumptions of a single dimensioned constant in each limit, and of no dimensionless constants that differ much from unity.
\par
Such ``mechanistic'' analogies might inspire construction of MOND theories (as in the case of the molecular-vortices auxiliary picture for the Maxwell equations).

And, add a footnote with a more formal explanation:
For a test particle of mass $m$, let $\B(\psi,G,\Qz)$ be a dimensionless ``criterion number'', whose value tells us whether we are in the DML or not (similar to $\hbar\n/kT$ in connection with the BB classical-quantum case, or $v/c$ in relativity). Here, $\psi$ represents some characteristic(s) of the particle.  Universality of free fall dictates that $\psi$ cannot depend on $m$. Then, since $\B$ is dimensionless, it is invariant under a change of the mass units, under which the values of $G$ and $\Qz$ do change. So, only the ratio $\qz=\Qz/G$, whose value does not change, can appear in $\B$. Furthermore, $\B(\psi,\qz)$ can depend only on $\psi$ that are either dimensionless or of the dimensions of $\qz$ and appear as $\psi/\qz$. The boundary is thus defined by $\qz$, but may depend on dimensionless characteristics, such as the type of orbit.

If all the test particles from which a system is made are always on one side of the boundary we may say that the whole system is on that side. In this case we may express the characteristic value of $\psi$ for a characteristic test particle in terms of the global properties of the system (such as mass and size), and get a global criterion for the system to be on that side of the boundary. For example, in the case of MOND where $\qz=\az$, such a system parameter may be $\s^2/R$, or $MG/R^2$, or $(M\azg)^{1/2}/R$. This is in the basis of the statement that a low-surface-density system: with $MG/R^2\ll\az$ is wholly in the DML.\footnote{Note that here the total mass $M$ does appear; this is not in conflict with the weak equivalence principle.}


\begin{thebibliography}{}
\bibitem{milgrom83}M. Milgrom, Astrophys. J. 270, 365 (1983).
\bibitem{milgrom83a}M. Milgrom, Astrophys. J. 270, 371 (1983).
\bibitem{fm12}B. Famaey and S. McGaugh, Living Rev. Relativity 15, 10 (2012).
\bibitem{milgrom14c}M. Milgrom, Scholarpedia, 9(6), 31410 (2014).
\bibitem{sn07}R. H. Sanders, and E. Noordermeer, Mon. Not. R. Astron. Soc. 379, 702 (2007).
\bibitem{humphrey11}P.J. Humphrey,{\it et al.}, Astrophys. J. 729, 53  (2011).
\bibitem{humphrey12}P.J. Humphrey, {\it et al.}, Astrophys. J. 755, 166 (2012).
\bibitem{milgrom12}M. Milgrom, Phys. Rev. Lett. 109, 131101 (2012).
\bibitem{brim13}F. Brimioulle, {\it et al.},  Mon. Not. R. Astron. Soc. 432, 1046 (2013).
\bibitem{milgrom13}M. Milgrom, Phys. Rev. Lett. 111, 041105 (2013).
\bibitem{milgrom09a}M. Milgrom, Astrophys. J. 698, 1630 (2009).
\bibitem{milgrom14}M. Milgrom, Mon. Not. R. Astron. Soc. 437, 2531 (2014).
\bibitem{loebman14}S.R. Loebman {\it et al.}, Astrophysical J. 794, 151 (2014).
\bibitem{milgrom94}M. Milgrom, Ann. Phys. 229, 384 (1994).
\bibitem{mb88}M. Milgrom and E. Braun,   Astrophys. J. 334, 130 (1988).
\bibitem{nv07}E. Noordermeer and M.A.W. Verheijen, Mon. Not. R. Astron. Soc. 381, 1463 (2007).
\bibitem{mcgaugh11a}S. S. McGaugh, Phys. Rev. Lett. 106, 121303 (2011).
\bibitem{mm13}S. McGaugh and M. Milgrom,  Astrophys. J. 766, 22 (2013).
\bibitem{mm13a}S. McGaugh and M. Milgrom, Astrophys. J. 775, 139 (2013).
\bibitem{sanders90}R.H. Sanders,  Astron. Astrophys. Rev., 2, 1 (1990).
\bibitem{tiret09}O. Tiret and F. Combes, Astron. Astrophys. 496, 659 (2009).
\bibitem{scarpa06}R. Scarpa, AIP Conf. Proc. 822, 253, arXiv:astro-ph/0601478 (2006).
\bibitem{milgrom99}M. Milgrom, Phys. Lett. A 253, 273 (1999).
\bibitem{planck15}P.A.R. Ade et al., Planck collaboration, arXiv:1502.01589 (2015).
\bibitem{milgrom09}M. Milgrom, Phys. Rev. D 80, 123536 (2009).
\bibitem{deffayet14}C. Deffayet, G. Esposito-Farese, and R.P. Woodard, Phys. Rev. D 90, 064038 (2014).
\bibitem{carroll04}S.M. Carroll and E.A. Lim, Phys. Rev. D 70, 123525 (2004).
\end{thebibliography}
\end{document}